\documentclass[preprint2]{aastex} % one-column, double-spaced document:

\usepackage{amsmath}
\usepackage{natbib}
\usepackage{graphicx}
\DeclareGraphicsExtensions{.pdf,.png,.jpg}
\pdfoutput=1

\shorttitle{MF and the SC}
\shortauthors{Upton \&Hathaway }

\begin{document}
\title{Effects of Meridional Flow Variations on Solar Cycles 23 and 24}
 
\author{Lisa Upton}
\affil{Department of Physics \&Astronomy, Vanderbilt University, Nashville, TN
37235 USA}
\affil{Center for Space Physics and Aeronomy Research, The University of Alabama in Huntsville,
Huntsville, AL 35899 USA}
\email{lisa.a.upton@vanderbilt.edu}
\email{lar0009@uah.edu}

\author{David H. Hathaway}
\affil{NASA Ames Research Center, Moffett Field, CA 94035}
\email{david.hathaway@nasa.gov}

\begin{abstract}

The faster meridional flow that preceded the solar cycle 23/24 minimum is thought to have led to weaker polar field strengths, producing the extended solar minimum and the unusually weak cycle 24. To determine the impact of meridional flow variations on the sunspot cycle, we have simulated the Sun's surface magnetic field evolution with our newly developed surface flux transport model. We investigate three different cases: a constant average meridional flow, the observed time-varying meridional flow, and a time-varying meridional flow in which the observed variations from the average have been doubled. Comparison of these simulations shows that the variations in the meridional flow over cycle 23 have a significant impact ($\sim$20\%) on the polar fields. However, the variations produced polar fields that were stronger than they would have been otherwise. We propose that the primary cause of the extended cycle 23/24 minimum and weak cycle 24 was the weakness of cycle 23 itself - with fewer sunspots, there was insufficient flux to build a big cycle. We also find that any polar counter-cells  in the meridional flow (equatorward flow at high latitudes) produce flux concentrations at mid-to-high latitudes that are not consistent with observations.

\end{abstract}

\keywords{Sun: dynamo, Sun: surface magnetism}

\section{INTRODUCTION}

Activity on the Sun is periodic with the average cycle lasting $\sim$11 years. \cite{Babcock61} explained this periodic behavior with a phenomenological solar dynamo model. In his model, the Sun's dipolar fields at solar cycle minimum are converted into toroidal fields within the Sun by differential rotation. These toroidal fields emerge at the surface in the form of tilted bipolar active regions. Surface flux transport on the Sun then carries the remnants of the active region flux to the poles where it cancels the original dipole and creates a new poloidal field with the opposite polarity. The strength of this new poloidal field (at solar minimum) is the seed to the next solar cycle. This theory is supported by findings \citep{Schatten_etal78, Svalgaard_etal05, MuozJaramillo_etal12, SvalgaardKamide13} that the strength of the Sun's polar fields during solar minimum is well correlated with (and can be used to predict) the amplitude of the following cycle. 

The polar field strengths have been measured directly by the Wilcox Solar Observatory for the last three solar cycles. The polar fields during the last solar minimum (cycle 23/24 minimum) were the weakest to date; they were $\sim$half the strength of the prior two cycles. This minimum proved to be exceptionally long and was followed by a cycle (cycle 24, the current solar cycle) that is developing into the weakest solar cycle in the last century. These unusual solar conditions may provide insight as to how magnetic flux transport at the surface regulates the polar field evolution and thus the solar activity cycle.

Surface flux transport on the Sun is achieved via differential rotation, meridional circulation, and the turbulent motions of convection. While the characteristics of the convective motions and differential rotation are fairly constant, the meridional flow varies in two fundamental ways: over the course of a solar cycle and from one cycle to the next \citep{Komm_etal93B, HathawayRightmire10, HathawayRightmire11}. The meridional flow is faster at solar cycle minimum and slower at maximum. Furthermore, the meridional flow speeds that preceded the cycle 23/24 minimum were $\sim$20\% faster than the meridional flow speeds that preceded the prior minimum. This faster meridional flow may have led to the weaker polar field strengths of cycle 23/24 minimum and thus the subsequent extended solar minimum and a weak cycle 24 \citep[see e.g.][]{SchrijverLiu08, Wang_etal09, HathawayRightmire10, HathawayRightmire11}. 

Surface Flux Transport models have long been used to characterize the dynamics of the Sun's magnetic fields. These models \citep{DeVore_etal84, Wang_etal89, vanBallegooijen_etal98, SchrijverTitle01, Baumann_etal04, Jiang_etal10} look exclusively at how magnetic flux moves at the surface, i.e. as a function of latitude and longitude. Surface flux transport evolves the magnetic fields with a given differential rotation, meridional flow, and supergranular diffusion. Most models have been highly parameterized, in particular with respect to the meridional flow and diffusion. The adopted meridional flow profiles (sharply peaked at low latitudes, stopping short of the poles, exaggerated variations around active regions) deviate substantially from the observed profiles.  Additionally, these models have typically neglected the variability in the meridional flow all together. Furthermore, virtually all previous models have parameterized the turbulent convection by a diffusivity with widely varying values from model to model.

Recently, \cite{Jiang_etal10} showed that perturbations on the meridional flow in the form of superimposed inflows into active regions (a strengthening of the flow on the equatorward side of the active regions and a weakening of the flow on the poleward side of the active regions) could modulate the strength of the polar fields, and thus the amplitude of the next cycle. \cite{CameronSchussler12} went on to suggest that these inflows were strong enough to reduce the tilt angle of the active regions and cause cross-equator flows that would enhance the cancellation of leading polarity flux across the equator. \cite{Yeates14} attempted to use the inflows suggested by \cite{CameronSchussler12}, but found that this led to a poorer correlation with the observations. This may be explained by the fact that our meridional flow observations indicate that the inflows into active regions are much weaker than suggested by \cite{CameronSchussler12}. We find that the presence of active regions causes the peak of the meridional flow to drop and move to lower latitudes --- giving a relatively faster poleward flow near the equator and a slower poleward flow at higher latitudes. Furthermore, no significant cross-equator flow is seen in the observations. 

We have developed a new surface magnetic flux transport model that advects the magnetic flux emerging in active regions using the observed near-surface flows including evolving convection cells that do not require any parameterizations or free parameters \citep{UptonHathaway14}. Here we use this model with the observed meridional flow variations to investigate the importance of these variations on the polar fields produced by cycle 23 --- the ultimate cause of the weak cycle 24 and exceptional cycle 23/24 minimum.

\section{THE SURFACE FLUX TRANSPORT MODEL}

We have created a surface flux transport model to simulate the dynamics of magnetic fields over the entire surface of the Sun. The basis of this flux transport model is the advection equation:
\begin{align}
\frac{\partial B_{r}}{\partial t} + \nabla \cdot (uB_{r}) &= S(\lambda,\phi,t)
\label{eq:advect} 
\end{align}
where $B_{r}$ is the radial magnetic flux, $u$ is the horizontal velocity vector (which includes the observed axisymmetric flows and the nonaxisymmetric convective flows), and S is an active region magnetic source term as a function of latitude ($\lambda$), longitude ($\phi$), and time ($t$).

This purely advective model is supported by both theory and observation. The Sun's magnetic field is carried to the boundaries of the convective structures (granules and supergranules) by flows within those convective structures. The field becomes concentrated in the downdrafts as small magnetic elements with radial (vertical) field. These weak magnetic elements are then transported like passive scalars (corks). This has been found in numerous numerical simulations of magneto-convection \citep[c.f.][]{Vogler_etal05} and is borne out in high time- and space-resolution observations of the Sun \citep{Simon_etal88, Roudier_etal09}. 

\begin{figure}[ht!]  
%\centerline{\includegraphics[width=1.0\columnwidth]{fig1.eps}}
\centerline{\includegraphics[ width=1.0\columnwidth ]{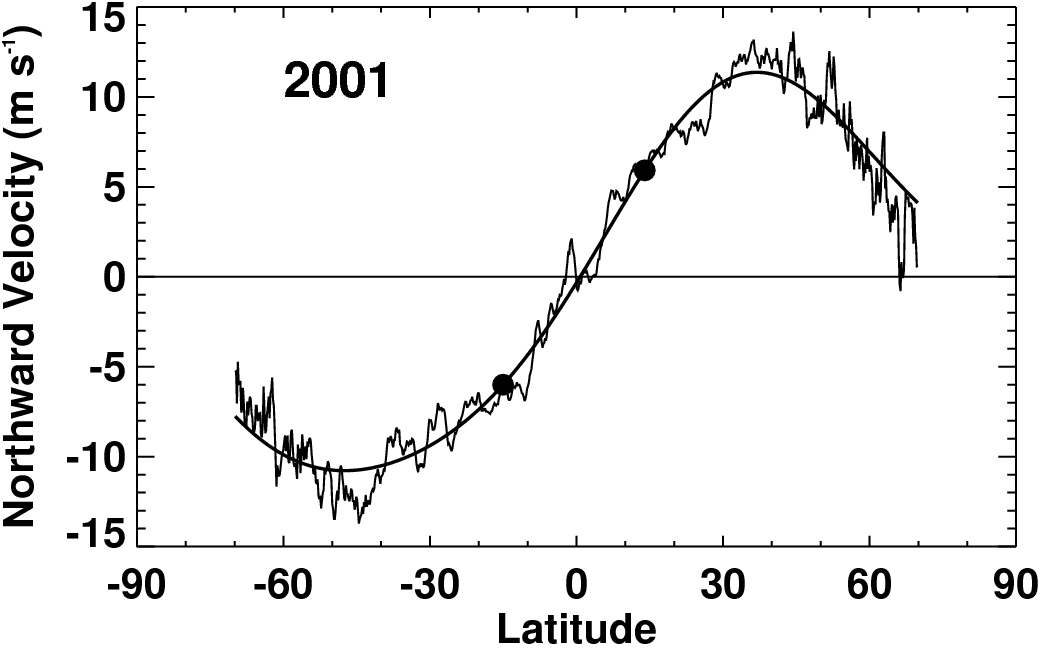}}
\caption[Legendre Polynomial Fits]{Legendre polynomial fit to the meridional flow in 2001.
The measured meridional flow profile averaged over calendar year 2001 is shown with the thin line.
The polynomial fit is shown with the thick smooth line. The filled black circles mark the active region centroid positions for that year.
It can be seen that even at cycle maximum that the polynomial fit is a good fit to the data.
Any additional superimposed flows must be relatively weak (1-2 m s$^{-1}$).  
}
\label{fig:MeridionalFlowProfile2001}
\end{figure}

We have measured the axisymmetric flows (meridional flow and differential rotation) averaged over 27-day rotations of the Sun by using feature tracking \citep{HathawayRightmire10, HathawayRightmire11, RightmireUpton_etal12} on full disk images of the Sun's magnetic field obtained from space by the \emph{Solar and Heliospheric Observatory} Michelson Doppler Imager (MDI) \citep{Scherrer_etal95} and by the \emph{Solar Dynamics Observatory} Helioseismic and Magnetic Imager (HMI) \citep{Scherrer_etal12}. These axisymmetric flow profiles were fit with Legendre polynomials up to 5th order in sin($\lambda$). The polynomial coefficients were smoothed (in time) using a tapered Gaussian with a full width at half maximum of 13 rotations. Figure \ref{fig:MeridionalFlowProfile2001} shows that these polynomial fits fully capture the structure of the flow profiles. These smoothed coefficients were used to update the axisymmetric flow component of the vector velocities for each rotation, thereby including the solar cycle variations inherent in these flows. 

The convective flows on the Sun, i.e., granules and supergranules, act not only to disperse magnetic flux, but also to concentrate flux into the magnetic elements of the magnetic network. \cite{Schrijver_etal96} showed that estimates of the diffusivity coefficient depend on the temporal and spatial scales, i.e. the effects of the convective flows cannot be captured by a single diffusivity alone. They state that ``...despite the apparent success of the description of flux dispersal as a diffusion process, there are several notable inconsistencies between that model and observations.'' Rather than employing a diffusivity with a Laplacian operator, we have chosen to use a supergranule simulation to explicitly model the convective motions. While the simulation is named for the dominant feature in spectrum, the supergranules, the simulation includes flow velocities for convective structures on all resolved scales (in this case, down to cells 6000 km in diameter).

The convective flows have been modeled using vector spherical harmonics as described by \cite{Hathaway_etal10} to simulate the convective motions on the Sun.  A spectrum of complex spherical harmonic coefficients was used to create convective flows that reproduce the observed spectral characteristics. The spectral coefficients were evolved at each time step to give the cells finite lifetimes and to make them move with the observed differential rotation and meridional flow. The convection cells have lifetimes that are proportional to their size, e.g. granules with velocities of 3000 m s$^{-1}$, diameters of 1 Mm, have lifetimes of $\sim$10 minutes and supergranules  with velocities of 500 m s$^{-1}$, diameters of 30 Mm, have lifetimes of $\sim$1 day. The evolving spectral coefficients are used to create a set of vector velocities that realistically simulate the flows observed on the Sun. For the simulations done here, vector velocities were created for the full Sun at 1024 pixels in longitude, 512 pixels in latitude, and at a 15 minute cadence.
These evolving flows automatically give different ``diffusivities'' at different spatial and temporal scales.

Outside of active regions, the magnetic fields are weak, the plasma beta is high, and the field elements are carried by the plasma flows. Inside active regions, the plasma beta is low so the flows themselves are modified (quenched) by the strong magnetic fields. To account for this, the supergranule flow velocities are reduced where the magnetic field was strong. While this model allows us to easily quench the velocities in active regions, this is not easily done with a diffusion coefficient. This aspect of the flux transport has not typically been captured in previous models.

The advection equation was solved with explicit finite differencing (first order in time and second order in space) to produce magnetic flux maps of the entire Sun at a cadence of 15 minutes. These \emph{synchronic} maps represent the Sun's magnetic field over the entire surface at a moment in time. The high convective velocities and high spatial resolution in the model can produce Gibbs phenomenon, i.e., ringing artifacts at sharp edges that can cause the solution to overshoot/undershoot in adjacent pixels. To stabilize the numerical integrations and mitigate this effect, a diffusion term is added so that Equation 1 becomes:
\begin{align}
\frac{\partial B_{r}}{\partial t} +  \nabla \cdot (uB_{r}) &= S(\lambda,\phi,t) + \eta \nabla^{2}B_{r} 
\end{align}
where $\eta$ is a diffusivity. We note that this diffusivity term was strictly for numerical stability. Unlike previous surface flux transport models, the addition of this term has little effect on the flux transport. The convective motions produced explicit random walks for the magnetic elements in this model. 

\begin{figure*}[ht!]
%\epsscale{2}
%\plotone{fig2.eps}
%\centering
%\includegraphics[ width=1.0\textwidth]{fig2.png}
\centerline{\includegraphics[ width=.9\textwidth]{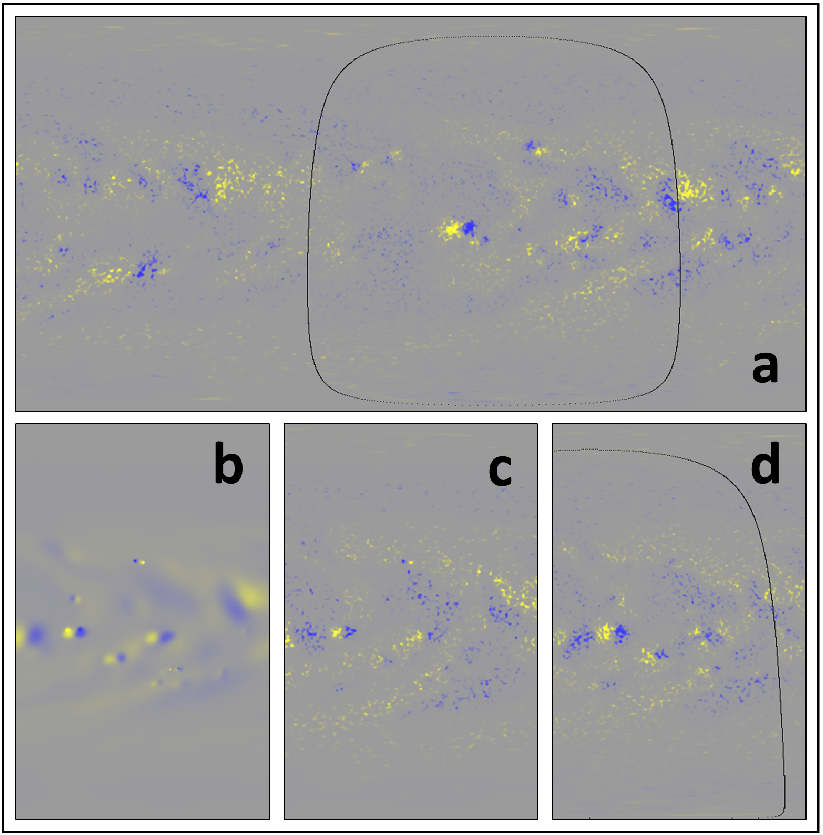}}
\caption{Flux transport method comparisons. A starting map (a) and a set of maps from one rotation of the Sun later: (b) a map from traditional surface flux transport that uses a diffusivity of 250 km$^2$ s$^{-1}$, (c) a map from our surface flux transport that uses the supergranule simulation, and (d) a map of the observed magnetic field.  It is clear from this figure that surface flux transport with the supergranule simulation is more realistic.
}
\label{fig:DiffusionTest}
\end{figure*}

The supergranule simulation far surpasses the realism that can be provided by a diffusion coefficient alone, making it both appropriate and effective for use in surface flux transport models. Figure \ref{fig:DiffusionTest} illustrates the improvements that this supergranule simulation provides. The figure compares surface flux transport that uses a diffusivity coefficient of 250 km$^2$ s$^{-1}$, surface flux transport that uses the supergranule simulation, and surface flux transport as observed on the Sun. The surface flux transport that uses the supergranule simulation provides the closest match to surface flux transport observed on the Sun, in particular the formation of the magnetic network.

Ideally, we would like detailed information on the active regions sources - location and magnetic flux of every sunspot at least once per day. The National Oceanic and Atmospheric Administration (NOAA) Solar Region Summaries provide information about the total sunspot area, location, and longitudinal extent of all the sunspot groups that have been observed since 1977 (cycles 21-24). Active regions emerge and grow over several days or weeks. Flux is constantly emerging and spreading out across the Sun's surface. While previous models have represented active region sources by the addition of a single dipole source on the day of maximum sunspot group area \citep{Sheeley_etal85, Wang_etal89}, we choose to add flux daily as the sunspot group grows. If a sunspot group's area is larger than its previous maximum, then we add the additional flux (in the form of a Gaussian spot pair) centered on the daily position given by the NOAA Solar Region Summaries. 

The increase in magnetic flux was calculated as a function of the increase in reported sunspot group area using the relationship described by \citet{Sheeley66} and by \citet{Mosher1977}:
\begin{align}
\Phi(A)  &= 7.0\times10^{19}A %\quad[\mbox{Mx}] 
\end{align}
\label{eq:fluxvsarea}
where $\Phi(A)$ is the magnetic flux in Maxwells and $A$ is the total sunspot area in units of micro hemispheres (1 $\mu$Hem = $3\times10^{16}$ cm$^{2}$). NOAA also provides the longitudinal extent of each active region from the leading edge of the sunspot group to the trailing edge of the sunspot group. The centers of the Gaussian spot pairs were placed at longitudes on either side of the reported position at $\pm 0.3$ times the given longitudinal extent and at latitudes given by the longitudinal separations and the average Joy's Law tilt i.e., the angle between the bipolar spots with respect to lines of latitude is equal to one half of the latitude \citep{Hale_eta19, StenfloKosovichev12}. 

Ideally we would want to include the emergence, and further development, of active region on the far side of the Sun. Unfortunately, detailed active region information for the far side of the Sun during the time interval we model does not exist. However, the flux we do miss on the far side is somewhat offset by the fact that we do include active regions as soon as they rotate onto the near side and some of these active regions were already added in during their previous near side disk passage.

\section{SIMULATIONS}

Our surface flux transport model was first used to create what we call a baseline data set. The baseline is created by allowing the model to assimilate (i.e. continually add in observed data weighted by it's noise level) magnetic data from magnetograms at all available longitudes and latitudes. The data assimilation process provides the closest contact with observations by correcting for any differences between data and model. In regions where data were recently assimilated (the outlined area in Fig. 2a), the baseline is virtually identical to the observations (including an annual signal due to instrumental effects). The baseline was used as a metric for comparison with the simulations in which the meridional flow varied. More detail about the baseline can be found in \cite{UptonHathaway14}.

\begin{figure}[ht!]  
%\centerline{\includegraphics[width=.85\columnwidth]{fig3.eps}}
\centerline{\includegraphics[ width=.9\columnwidth]{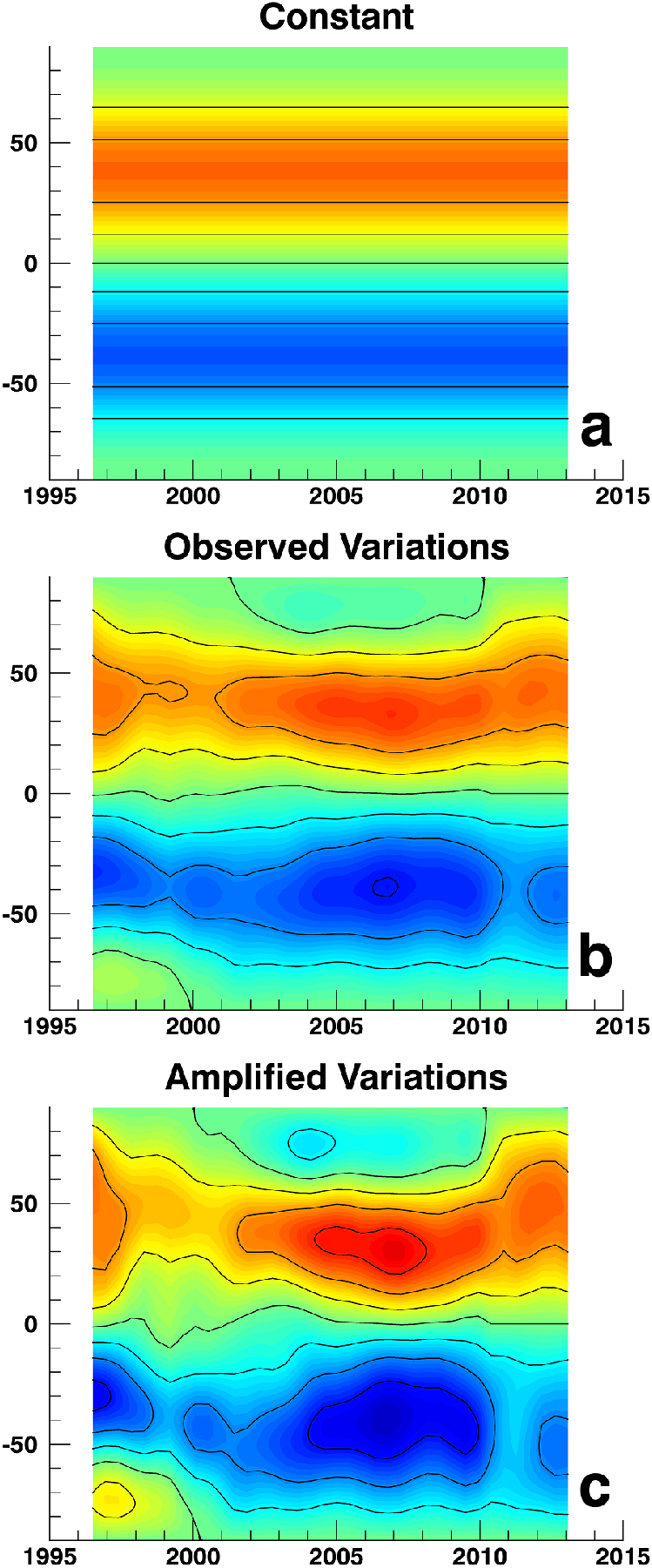}}
\caption[Variations in the Meridional Flow]{The meridional flows as functions of latitude and time. (a) The constant North-South antisymmetric meridional flow used in Sim 1. (b) The observed meridional flow used in Sim 2. (c) The meridional flow with amplified variations used in Sim 3.  Red is northward flow and blue is southward flow. For reference the contours (black lines) show 0, $\pm$5, $\pm$10, and $\pm$15 m s$^{-1}$.  
}
\label{fig:PhDMFPlots}
\end{figure}

The model was then used in three simulations to determine the impact of the observed meridional flow variations on the sunspot cycle. Each simulation was run from January 1997 through July 2013. All three models used the same average, constant, and North-South symmetric differential rotation, given in terms of latitude ($\lambda$) by
\begin{align}
v_{\phi}(\lambda)&= [A +B\sin^{2}(\lambda) +C\sin^{4}(\lambda)]\cos(\lambda)
\end{align}
with
\begin{equation}
\begin{split}
A&= 39 \text{ m s}^{-1}\\ 
B&= -244 \text{ m s}^{-1}\\ 
C&= -374 \text{ m s}^{-1} 
\end{split}
\end{equation}
The supergranule realizations were virtually identical for all three simulations. The size, lifetimes, and initial locations of the cells were the same. However the latitudinal drift of the convective cells were modulated by the meridional flow applied to each simulation. The meridional flow for each simulation was updated at six month intervals. Figure \ref{fig:PhDMFPlots} shows the meridional flow profiles for all three simulations as functions of latitude and time. 

The first simulation (hereafter referred to as Sim 1) included an average, constant, North-South antisymmetric meridional flow given by
\begin{align}
v_{\theta}(\lambda)&= [D\sin +E\sin^{3}(\lambda) +F\sin^{5}(\lambda)]\cos(\lambda)
\end{align}
with
\begin{equation}
\begin{split}
D&= 24 \text{ m s}^{-1}\\ 
E&= 16 \text{ m s}^{-1}\\ 
F&= -37 \text{ m s}^{-1} 
\end{split}
\end{equation}
The second simulation (Sim 2) included a meridional flow with the observed North-South structure and variations in time. In the third simulation (Sim 3), the differences between the average meridional flow and the observed meridional flow were doubled to amplify the variations in the  meridional flow. The meridional flows in Sim 2 and Sim 3 included additional terms with even powers of $\sin(\lambda)$ to account for the observed North-South asymmetries - most notably the polar counter-cells observed with MDI. 

Figure \ref{fig:PhDMFPlots} shows that the observed meridional flow varied in four stages relative to the average meridional flow. At the start of the simulations (1997-1998) the observed flow was similar in strength to the average flow (but with a counter-cell in the south). Around the time of the maximum of cycle 23 (1999-2002) the observed flow was slower than the average flow. By 2003 the flow speed had increased to become faster than average and it remained faster until 2010. This was then followed by rapid decrease in flow speed and a return to average. Note that the flow speed was average at cycle 22/23 minimum and average again at the maximum of cycle 24. In general, the meridional flow was slow at both the start and at the maximum of cycle 23 when compared to similar phases of cycles 21, 22, and now 24.

\section{RESULTS}

\begin{figure}[ht!]  
%\centerline{\includegraphics[width=1.0\columnwidth]{fig4.eps}}
\includegraphics[ width=1.0\columnwidth]{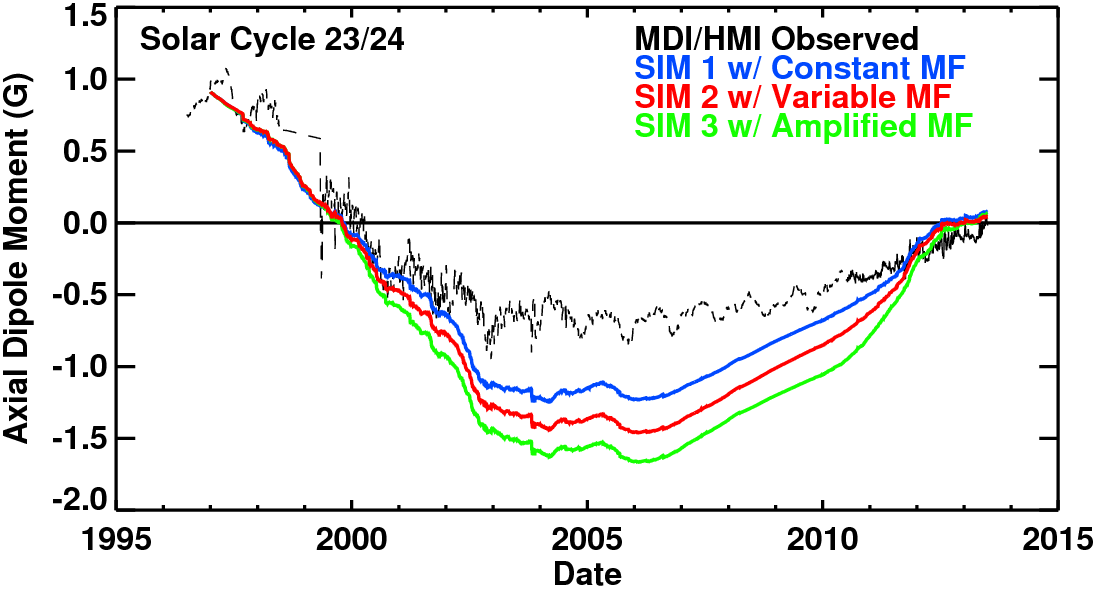}
\caption[Axial Dipole Simulations]{The evolution of the axial dipole moments with different meridional flow profiles. The baseline axial dipole moment is shown in black (dashed for MDI, solid for HMI). Sim 2 (red, with the observed variations in the meridional flow) produced an axial dipole moment at cycle 23/24 minimum (late 2008) that is approximately 20\% stronger than Sim 1 (blue, with the constant average meridional flow profile). Sim 3 (green, with the amplified variations) made the dipole another 20\% stronger than in Sim 2. 
}
\label{fig:MFEffectOnDipole}
\end{figure}

Each simulation produces synchronic maps (magnetic flux maps of the entire Sun) at a cadence of 15 minutes. These maps were used to calculate the axial magnetic dipole moment $B_{p}$ at each time step, where $B_{p}$ is given by
\begin{align}
B_{p}  &= \int_{0}^{2\pi} \int_{-\pi/2}^{\pi/2} B_{r}(\lambda,\phi) Y_{1}^{0}(\lambda,\phi) \cos\lambda d\lambda d\phi.%.\] 
\end{align}
The axial dipole moments measured for the baseline and each simulation are plotted in Figure \ref{fig:MFEffectOnDipole}.

A quick inspection of these axial dipole moments reveals that the dipole moments in the simulations reach amplitudes about twice as strong as the observed axial dipole moment. Given the fact that the relative positions and magnetic flux we added for the active region sources were based on simple statistical relationships, it is not at all surprising that the match is less than perfect. However, the systematic offset suggests that the active region sources that most influence the axial dipole moment (total flux and/or latitudinal separations) are systematically over-estimated. Remarkably (and despite this) the timing of the cycle 24 reversal (in 2012/2013) is accurate to within about a year. Using observed magnetic flux and its location in active regions might improve the model to the point that the axial dipole moment for an entire solar cycle is reproduced with the observed flows. 

Somewhat surprisingly, rather than producing weaker polar fields, the observed meridional flow variations produced a stronger axial dipole moment. Furthermore, this effects was more pronounced with the amplified meridional flow variations (Sim 3), indicating that the flow variations were indeed the source of the difference. 

In all three cases, the axial dipole is well matched for the first three years as the dipole steadily and rapidly reverses polarity during the rise to maximum of cycle 23. During this time, the meridional flow was similar in all three simulations and consequently had the same effect on the axial dipole moments.

This dipole reversal is caused by the steadily increasing active region emergence combined with Joy's Law, Sp\"orer's Law, and the poleward transport of the following polarity flux. (Joy's Law is the characteristic tilt of bipolar active regions such that the leading polarity is more equatorward than the following polarity and Sp\"orer's Law says that active region emergence progresses towards the equator over the course of the cycle.) The following polarity of newly emerging active regions cancels and then exceeds the leading polarity of the older active regions at progressively lower latitudes. The excess following polarity flux is transported to poles where it cancels the dipole field established at the previous minimum.

From 2000 to about 2004 the axial dipole moments in the three simulations follow different paths with the dipole moments getting stronger in Sim 2 and stronger yet in Sim 3. We attribute this divergence to the slowdown in the meridional flow in these simulations. By 2000 (cycle 23 maximum) the active regions cover their widest range of latitudes and many active regions emerge close to the equator. The slower meridional flow at this time allows more leading polarity flux to be transported across the equator by the convective motions. This leaves behind a greater excess of following polarity flux which, when transported poleward, increases the axial dipole moment. (Note that it takes a year or two for active region flux to be transported to the polar regions. This gives a similar lag in the response of the axial dipole moment to the variations in the meridional flow.)

After 2004 cycle 23 was in decline and the meridional flow speed increased substantially. With fewer active region sources and less cancellation across the equator it was more difficult to continue building up stronger polar fields. Furthermore, during solar minimum very little flux is emerging in the form of active regions.
Without flux to transport, the meridional flow has little to no effect. Instead, turbulent motions slowly
erode the flux at the poles and the axial dipole moment gradually decays in all three simulations.

From 2010 to 2012, the cycle 24 field reversal occurs with Sim 3 reversing the fastest and Sim 1 reversing the slowest. While we attribute this to the rapid slow down of the meridional flow around 2010, the active regions are emerging far from the equator and thus it is too early for any cross-equatorial flow. We conjecture that the slower meridional flow may simply allow for more cancellation in the active latitudes, resulting in less following polarity flux being transport to the poles, and thus the different reversal rates. We note that this is also precisely the time when the counter-cell in the north disappears. The sudden lack of a counter-cell in the north may cause more following polarity flux to be picked up by the meridional flow and carried to the poles, exacerbating the different reversal rates. After 2012 all three simulations have similar developments since the meridional flow is nearly average in all three.

\begin{figure*}[ht!]
%\epsscale{2}
%\plotone{fig5.eps}
\includegraphics[ width=\textwidth]{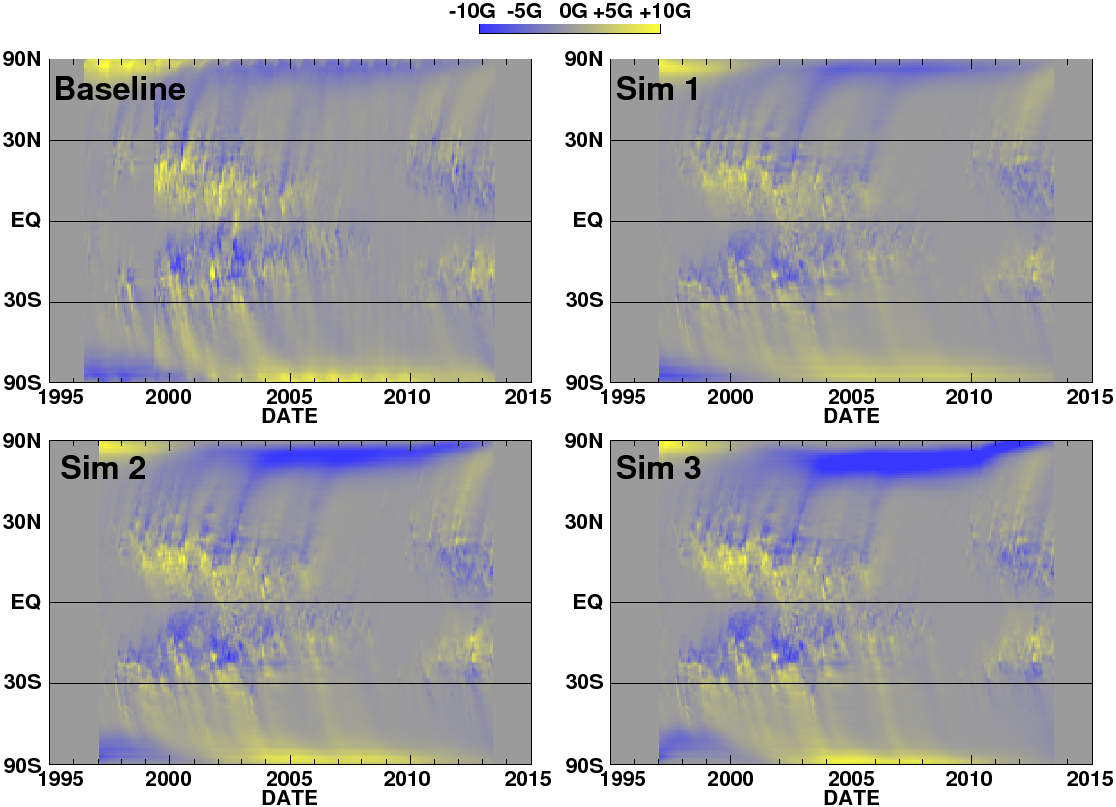}
%\centerline{\incbb=0 0 1116 807, h=1.0\columnwidth]{fig5.png}}
\caption{Magnetic Butterfly Diagrams. These magnetic butterfly diagram illustrate the evolution of magnetic flux in the baseline and in the three simulations. The baseline is shown on the top left. Sim 1 (top right) used a constant  north-south symmetric meridional flow. The observed meridional flow was used in Sim 2 (bottom left). Sim 3 (bottom right) used the observed meridional flow with variations from the average doubled.
}
\label{fig:magbflyPhDSimCombined}
\end{figure*}

To provide additional context and information, the magnetic butterfly diagrams of the baseline and each simulation are shown in Figure \ref{fig:magbflyPhDSimCombined}. 

A comparison of the simulated magnetic butterfly diagrams with the baseline illustrates the fidelity of this flux transport model. The simulated magnetic butterfly diagrams share details with the baseline in spite of the lack of detail in our active region sources (i.e., total flux, tilt, and longitudinal separation). The mottled pattern produced by active region emergence (i.e. the butterfly `wings') are well reproduced, with distinct features often visible in both the baseline and the simulations. Poleward streams of both polarities are also observed, often with a one-to-one correspondence between the baseline and the simulations. 

The biggest discrepancies between the baseline and the simulations appear at the poles, and in the north in particular. First of all, the amplitude is too strong, further indicating that the sources are being over-estimated. Secondly, we note that the polar concentrations occur at lower latitudes for Sim 2 and Sim 3 when counter-cells are present in the meridional flow (in the south prior to 2000 and in the north from 2000 to 2010). The fact that this is not observed in the baseline, nor with the HMI data, suggests that these polar counter-cells are not real. The presence of counter-cells in the meridional flow as measured by MDI could be caused by unaccounted for changes in the geometry (e.g., elliptical distortion) of the MDI images.

\section{CONCLUSIONS}

Our innovative surface flux transport model features several advantages over most previous surface flux transport models:
\begin{enumerate}
  \item This model advects the flux with simulated convective motions, rather than invoking a diffusion coefficient to account for the turbulent transport.
  \item This model incorporates the observed meridional flows, whereas previous models have used meridional profiles that deviate significantly from the observations. 
  \item In the Sun, the convective flows are affected by strong magnetic fields (i.e., the flows are quenched in active regions). While this effect is difficult to reproduce with the diffusivity used in prior models, this effect is captured in our model by actually quenching the convective flows where the field is strong.
	\item While previous models have added a single magnetic flux source on the day of maximum sunspot group area to represent sunspots, our model adds flux daily as it emerges.
\end{enumerate}
These advantages make this model the most realistic surface flux transport model to date.  

The simulations presented in this paper show that the variations in the meridional flow over cycle 23 had a significant impact (at least $\sim$20\%) on the polar fields. The variable meridional flow produced a stronger axial dipole moment than was produced with a constant meridional flow. We are confident that this is a robust result that would also be obtained with other surface flux transport models using our meridional flow and its variations. The meridional flow variations (in particular the very slow flow at cycle 23 maximum) appear to have kept the polar fields from becoming even weaker yet. The weak polar fields at cycle 23/24 minimum caused the extended minimum and then went on to produce the weak cycle 24. If the polar fields were even weaker they might have led to a grand, Maunder-type, minimum. This suggests that variations in the meridional flow may provide a possible feedback mechanism for regulating the solar cycle and possibly recovering from a Maunder-type minimum \citep[see e.g.][]{Jiang_etal10, CameronSchussler12}. 

There are two possible ways for producing the observed weak polar fields that modulate the solar cycle: 1) via flux transport or 2) via the active region sources. There is no credible evidence that the diffusion process changes substantially, but there is significant variation in the meridional transport. This experiment suggests that the actual cause of the weak polar fields at the end of cycle 23 was not the meridional flow variations. By eliminating the meridional flow as a cause of the weak polar fields, the active region sources must be the culprit.

Active region sources can modulate the polar field strengths in two ways: 1) via the amount of flux emerging at the surface (i.e., the amplitude of the sunspot cycle) or 2) via the orientation of the active regions (i.e., the latitudinal separation of the opposing polarities). We propose that the most simple and reasonable explanation for the weak polar fields at the end of cycle 23 is the emergence of fewer active region sources. Cycle 23 had a peak sunspot number of $\sim$120 - much smaller that cycle 21 and cycle 22 (which had peaks of $\sim$160). With fewer active region sources to reverse the strong polar fields at cycle 22/23 minimum (in 1996), there was insufficient flux to rebuild a strong polar field. 

This conclusion, while simple and supported by observations of cycle 23, is contrary to conclusions by others. In particular, \citet{Jiang_eta13}  concluded that the weakness of the polar fields at the end of cycle 23 could have been produced by an increase in the meridional flow by 55\%, or by a decrease in active region tilt by 28\% (but with a 1.5 year delay in the polar field reversal relative to what was observed). While they also found that a 40\% reduction in the sunspot number could account for the weaker polar fields, they discounted this explanation because it did not produce enough open flux. 

There are significant differences between their study and ours, especially in the details of the flux transport itself. They used an idealized meridional flow profile significantly different from the profile we observe (no meridional flow at all above 75 degrees latitude). Additionally, their meridional flow did not include the observed systematic variations (amplitude and latitudinal structure) over the course of the cycle. They parameterized the advection of the magnetic field by the convective motions by a diffusivity (250 km$^2$ s$^{-1}$) and a Laplacian operator rather than explicitly including those flows. They also used idealized active region sources instead of actual active region sizes and locations. Different values for the meridional flow, the diffusivity, or for the active region sizes and locations would impact their conclusions.

Their conclusions were based on comparisons to their reference model - which produced strong polar fields at the end of cycle 23 (even stronger than those at the end of the much larger previous two cycles). This is inconsistent with the polar fields observed on the Sun for Cycle 23, which were about half as strong as for the prior two cycles. The production of these strong polar fields in their reference model can be traced to two significant differences in their active region sources. First of all, they included changes in Joy's law tilt, with their reference case having 10\% more tilt in cycle 23 than in the two preceding cycles. Secondly, they did not use actual data for the sizes and locations of active regions in these cycles. Instead they used the monthly sunspot number as a parameter in a set of empirical relations that generate artificial sizes and locations for active regions. An important difference, evident from comparing butterfly diagrams of active region locations, is the continued emergence in their data of active regions in the north after 2005. This is notably absent in actual cycle 23 data.

While the active region tilt and meridional flow no doubt play a role in modulating the polar fields, as suggested by \citet{Jiang_eta13}, the importance of the amplitude of the cycle may have been masked by other details in their flux transport model. In this paper, we have performed a fairly simple scientific experiment in which the only variable was the meridional flow. We found that the observed variations in the meridional flow produced a stronger axial dipole moment relative to that produced by the average meridional flow. Based on these results, we conclude that the primary reason the polar fields were weak at the end of cycle 23 is because the active region sources were weak.

Comparison of the magnetic butterfly diagrams showed that polar counter-cells in the meridional flow produce magnetic flux concentrations that are offset to lower latitudes from the poles. This effect was previously noted by \citet{Jiang_etal09}. In the baseline, however, the flux is concentrated precisely at the poles. This is evidence that the polar counter-cells observed in the meridional flow measurements are likely to be artifacts. Furthermore, the full magnetic butterfly diagram history does not contain any offset polar magnetic field concentrations. This indicates that the meridional flow has not had any polar counter-cells during this time period, i.e., from 1973 to present. 

The presence of poleward streams of \emph{leading} polarity flux provides insight into the nature of these streams. In the past, these poleward streams have been attributed to active regions with large variations from Joy's law tilt. However, the current version of the flux transport model uses the average Joy's law tilt for each active region source without any variations. Since these streams can be reproduced without the scatter in the tilt of bipolar active regions, we propose an alternate explanation.

Normally, new active regions emerge slightly equatorward of the older decaying active regions. This causes the following polarity of the new active regions to emerge at similar latitudes to the leading polarity flux in old active regions. The excess following polarity flux is then transported to the poles forming a following polarity poleward stream, while the new leading polarity flux is left to be canceled by future following polarity emergence. However, if a substantial gap or a sudden equatorward shift in active region emergence occurs, there would be nothing to cancel with the leading polarity flux. This leading polarity flux would then be transported (after the following polarity flux) to the poles in the form of a leading polarity poleward stream. Alternatively, the appearance of an unusually strong bipolar active region (or progressively weaker active region emergence occurring in the declining phase of the solar cycle) may produce the effect. While active regions with substantial deviations from Joy's law probably do contribute to the production of these streams, this is clearly not the only source.    

The fact that all three simulations produced axial dipole moments and polar fields much stronger than observed suggests that our active region sources need revision. Active region flux versus sunspot group area is a crucial component in our model. We suspect that this relationship may be in error. \cite{Sheeley66} and \cite{Mosher1977} found a linear relationship between these parameters; however this was based on very few data points. \cite{Dikpati_etal06} found a linear relationship with a nonzero intercept that gives about 7-times as much flux, but argue that only 1/7$^{th}$ survives immediate cancellation with nearby opposite polarity flux. Further investigation of this this relationship using MDI and HMI data may be needed to improve the predictive capability of this flux transport model. Our surface flux transport model (which can use active region databases to replicate magnetic field emergence) has been shown to accurately predict the polar field evolution for 3-5 years \citep{UptonHathaway14}. Improvements to the active region flux versus area relationship may extend the predictive capability of the model by years. Remarkably, and despite this impairment, the timing of the polar field reversals were accurate to within a year.

\acknowledgements
The authors were supported by a grant from the NASA Living with a Star Program to Marshall Space Flight Center. The HMI data used are courtesy of the NASA/SDO and the HMI science team. The SOHO/MDI project was supported by NASA grant NAG5-10483 to Stanford University. SOHO is a project of international cooperation between ESA and NASA.

\clearpage

\bibliographystyle{apj}
\bibliography{MyBib}

\end{document}